# An Efficient Interference Aware Partially Overlapping Channel Assignment and Routing in Wireless Mesh Networks

Sarasvathi V[1], N.Ch.S.N. Iyengar[2], Snehanshu Saha[3]

[1]Department of Computer Science and Engineering, PESIT Bangalore South Campus, Bangalore-560 100, India.
[3]School of Computing Science & Engineering, VIT University, Vellore-632014, Tamilnadu, India.
[3]Department of Computer Science and Engineering, PESIT Bangalore South Campus, Bangalore-560 100, India.

sarasvathiram@gmail.com[1], nchsniyengar48@gmail.com[2], snehangshu.saha@gmail.com[3]

*Abstract:* In recent years, multi-channel multi-radio wireless mesh networks are considered a reliable and cost effective way for internet access in wide area. A major research challenge in this network is, selecting a least interference channel from the available channels, efficiently assigning a radio to the selected channel, and routing packets through the least interference path. Many algorithms and methods have been developed for channel assignment to maximize the network throughput using orthogonal channels. Recent research and test-bed experiments have proved that POC (Partially Overlapped Channels) based channel assignment allows significantly more flexibility in wireless spectrum sharing. In this paper, first we represent the channel assignment as a graph edge coloring problem using POC. The signal-to-noise plus interference ratio is measured to avoid interference from neighbouring transmissions, when a channel is assigned to the link. Second we propose a new routing metric called signal-to-noise plus interference ratio (SINR) value which measures interference in each link and routing algorithm works based on the interference information. The simulation results show that the channel assignment and interference aware routing algorithm, proposed in this paper, improves the network throughput and performance.

## 1. Introduction

As the technology nurtures the wide range of different gadgets, the number of clients utilizing Wireless Mesh Networks (WMN) for Internet access has also been increased to a large extent. This is because of the diverse applications provided by WMNs. A few of these samples are: military applications, Municipal Wireless Mesh Networks, public safety agencies and mining [1-3]. The network would work fine until the current frequency distribution is not disturbed in anyways. But it is also the necessity of the technology to host all the clients coming into the scope of a network, and maintaining the transparency at the same time. Interference plays a major role when the network comes into the picture of frequency sharing among many different clients. In order to avoid interference and fair bandwidth distribution [14], the number of clients accessing the network would have to be limited, and then there is no point in praising the technology. It is the technology which needs to be evolved at the right phase to overcome the interference, and provide the maximum utility to the network clients as well as taking care of unpredictable network scaling.

While the IEEE 802.11 standards, which were formed during 1990s, uses 3 non-overlapping channels of the frequency spectrum, and the remaining 8 channels are left unused. To suit the current network fan-out scenario, it is a must to utilize the available bandwidth effectively and efficiently. Using the multi-channel and multi–radio, it is possible to achieve the maximum performance, without bandwidth degradation. Multi-radio refers to a dedicated NIC assigned to each link on the mesh node and each link is assigned with a unique frequency (channel) for parallel data transmission.

In this paper, we work for the hybrid multi-channel and multi-radio Wireless Mesh Networks (MCMR-WMNs) that avoid interference to improve the network connectivity and enhance the throughput**.** Figure 1 depicts hybrid wireless mesh network which is based on 3 tiers of devices: 1) The lowest tier includes Wi-Fi, Wi-Max, Cellular networks, conventional clients and mobile nodes. 2) The second tier includes routers that relay packets between the user and Gateway. 3) The highest tier is Gateway. Routers with gateway capability are connected to the internet with wired connection.

Interference is a threat behind utilizing the network bandwidth proficiently and achieving the effective throughput. In MCMR-WMNs, the key challenge is the interference [13] of simultaneous data transmission that will worsen the network parameters and sequentially the capacity of the network. Hence we concern minimizing the interference, improving throughput, and scaling the network by effectively assigning the available channels to the respective radios and then finding optimal path to the destination.

We consider IEEE 802.11b/g based wireless mesh network operating in 2.4 GHz frequency band. The IEEE 802.11b/g standard has a total of 11 frequency channels available for transmission, of which 3 are orthogonal channels. (Figure 2**)** Each channel of the spectrum specifies the center frequency used by the transceiver and the AP within the range; Channel 1 uses 2.412 GHz and channel 2 uses 2.417 GHz. The Guard band that separates two center frequencies is only 5MHz and the signal approximately make up to 22 MHz of the available frequency spectrum. This obviously leads to overlapping the adjacent channels which, when used for data transmission, causes interference and leads to collision of data. To use multiple channels of the available bandwidth, only one out of 5 consecutive channels can be employed simultaneously. This proves that only 3 orthogonal channels are available, and can be used with the existing channel utilization techniques to prevent interference**.** A solution defined for this problem is the Partially Overlapped Channel (POC) assignment [12], where the channels can be selected with 2 other channels apart, and the two simultaneous transmissions are 10m apart. This also gives the same throughput in analysis [5], when compared with the existing non-overlapping channel assignment technique. In a dense wireless mesh networks, using all the available channels that are assigned properly with spatial separation would result in a higher performance than the orthogonal channels.

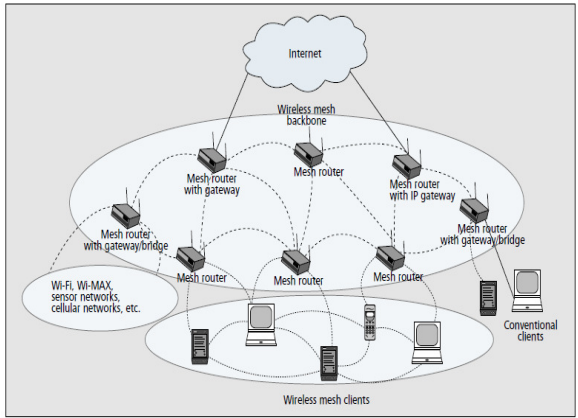

**Figure 1:** Hybrid Wireless Mesh networks

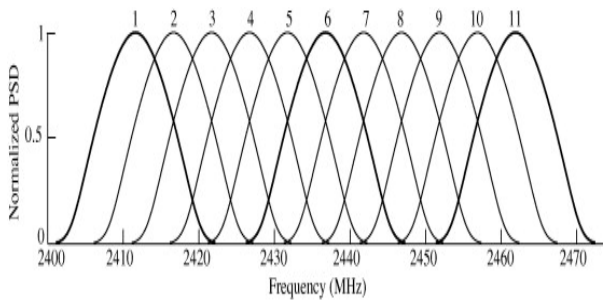

**Figure 2:** Available eleven partially overlapped channels in 802.11b frequency band

The following sections of the paper are organized as follows. In Section II, we describe previous work in centralized and distributed channel assignment algorithms. Section III, elaborates the types of interference and its consequences on network parameters. The channel assignment with Partially Overlapped Channel model using edge coloring algorithms and Routing algorithm using SINR metric are explained in section IV. Section V elaborates simulation and performance evaluation. Finally we conclude our paper with summary of our findings and the future plan in section VI.

## 2. Related Study

There are number of channel assignment problems [15] that have been proposed to reduce interference and increase throughput. Most of the channel assignment algorithms are combined with either routing or congestion control techniques. A centralized load-aware channel assignment was proposed by Raniwala *et al* [6] which uses central controller. All information about traffic and routing paths are informed to the central controller; whereas in the model we propose, the gateway does not require any such knowledge for channel assignment except the set of active links. This centralized load aware [11] channel assignment is combined with routing algorithms.

JOCAC algorithms [7] assign channels, based on average congestion price on links. JOCAC is described as a network utility maximization problem and it can be implemented either in a distributed or centralized manner. In [4, 5], Mishra proved that the capacity of the network can be increased when both non-overlapped and POC are used for channel assignment. Authors demonstrated that good spatial re-use of same POC gives better performance. MMAC protocol [8] was proposed by J. So et al for controlling multi-channel assignment using a single transceiver. The MMAC protocol uses orthogonal channels and it uses two channels for data packets and one for control packets. Also power saving mechanism is integrated with MMAC protocol for efficient energy saving. In [9], authors used Load-Aware CAEPO, which uses partially overlapped channels along with orthogonal channels. They proposed grouping algorithm which selects one group leader within its interference range. Each group leader selects the best channel from the channels available.

A good routing algorithm and routing metric mitigate and detect the interference experiences on the network. The traditional hop count metric may not give a good result in WMN, because it doesn't consider wireless link quality metrics such as packet loss rate, interference and load; it also gives equal consideration to all the links.

The Expected transmission count (ETX) metric of a link is calculated based on the number of attempts required to deliver packets successfully over a given link. ETX is computed for each link using delivery ratios of the link in both directions [16].

$$\text{ETX} = \frac{1}{d_f \times d_r} \quad (1)$$

Where $d_f$ and $d_r$ are the forward and reverse delivery ratios. The $d_f$ is the probability measure of a data packet successfully reaches at the destination. The $d_r$ is the measured probability that the ACK of the packet is reached successfully. ETX does not take load, interference and data rate of link. So, it is suitable only for the single channel multi hop network and it doesn't suit for multi-channel multi hop network.

Expected transmission time (ETT) is better than ETX, as it considers data rate (bandwidth) of link into account. ETT of a link is defined as a bandwidth adjusted ETX [17]. Let S denote the packet size and B the bandwidth of the link. Then ETT is computed by

$$\text{ETT} = \text{ETX} * \frac{S}{B} \quad (2)$$

The ETT does not explicitly consider the impact of traffic coming from other nodes. The disadvantage of ETT is that the contending traffic raises the loss rate of packet due to congestion, and it also decreases the available bandwidth.

The weighted cumulative expected transmission time (WCETT) [17] is designed to ameliorate the ETT metric. WCETT considers channel diversity, so it can be used in multi radio mesh networks. The WCETT metric of a path p is defined as follows:

$$WCETT_p = (1-\alpha) \times \sum_{i \in p} ETT_i + \alpha \times \max_{1 \leq j \leq k} X_j \quad (3)$$

Where Xj is the sum of the Expected transmission time values of links that are on channel j. k is the number of orthogonal channels and α is a tunable parameter subject to 0 ≤ α ≤ 1. WCETT enhances the performance of multi radio and multi rate wireless networks, compared to other metrics such as, hop count, ETX and ETT. The main disadvantage of WCETT is that it does not take the inter-flow interference into an account.

Many types of link metrics have been proposed in WMN to minimize interference and load on the link. Each link metric has some advantage and limitation, and gives better results in particular environment. In our work, the first proposal is that the channel assignment problem using the graph edge coloring method and we propose a new routing metric called signal-to-noise plus interference ratio (SINR) value which measures interference in each link and then routing algorithm works based on the interference information.

## 3. Interference

Interference plays a vital role in WMNs, which will lead to undesirable consequences. Concurrent transmissions of nodes situated close to each other in WMN increases interference and reduce the network throughput. The interference degrades the capacity of the network. For a restricted-bandwidth continuous-time stochastic channel that may endure noise, Shannon-Hartley Theorem provides the channel capacity (in bps)

$$C = B \log (1 + S/N) \qquad (4)$$

Where B is the channel bandwidth (in Hertz) and S/N is the signal-to-noise ratio.

This theorem also helps to discern the different types of interferences. Consider four nodes X1, X2, Y1 and Y2. Let X1 and X2 be sending nodes and Y1 and Y2 be receiving nodes. All four nodes are situated within the interference range of each other.

### 3.1 Co-Channel Interference

Co-channel interference is generated when two different communicating pairs of nodes within the transmission range of each other that use the same channel simultaneously as shown in figure 3. Consider X1-Y1 and X2-Y2 are assigned to channel 6 and following are the sequence of events: Node X1 wants to send data to Y1. It senses the medium (Channel 6) and checks if it is busy or idle. In the event of the medium is busy, the node will wait for its transmission. If the medium is idle, it will start the data transmission. When X1 is sending data to Y1 and at the same time X2 also tries to send a data to Y2, applying the same medium sensing procedure. In this case, the medium (Channel 6) will be busy. Hence, X2 waits for a back-off period and keep attempting till the data transmission between X1 and Y1 is completed. When X2 identifies the medium as free, it proceeds in transmitting the data. The CSMA/CA Co-channel interference is less harmful compared to adjacent channel interference.

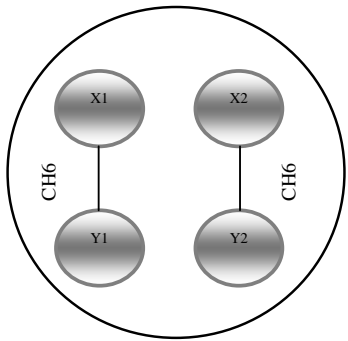

**Figure 3:** Co-Channel Interference

### 3.2 Self-Interference

Self-Interference is caused when two different nodes connected to a common node, assigned with same frequency as shown in figure 4. Consider X1 is sending packet to Y1 and Y2 simultaneously and the node X1 is equipped with two radios. If both the interfaces are assigned to channel 6 and X1 attempts to simultaneously transmit packets on both the NICs. In this case, the interference will be severe, irrespective of the distance of the receiving node whether it is located near or far. If mutually orthogonal channels are assigned to links in a single node, self-interference can be avoided.

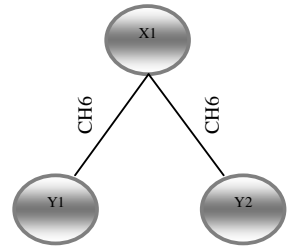

**Figure 4:** Self-Interference

### 3.3 Adjacent channel Interference

Adjacent channel interference occurs when the frequency of one transmission partially overlaps with the neighboring channel as shown in figure 5. Consider X1-Y1 and X2-Y2 are assigned to channels 6 and 8, respectively. X1 starts transmitting first, X2 will notice the medium as free in channel 8 and also begins to transmit its packets. Since the channel 6 and 8 share the frequency band, the receiving nodes Y1 and Y2 can't decode the packets correctly, causing a transmission error that severely decreases the network throughput.

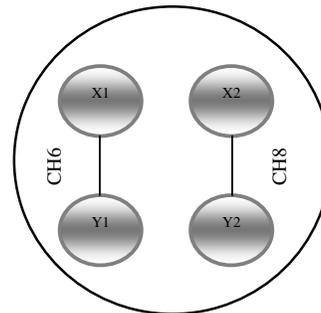

**Figure 5:** Adjacent Channel Interference

## 4. System model

The Figure 6 illustrates the relationship between the components used in our framework; it shows how the data flows among the components and the interaction between the Channel assignment and routing algorithms using SINR computation. The interference of each link is calculated during the channel assignment and this information is stored into the interference database. In our work, the channel assignment is completed before the routing algorithm is processed, and the channel assignment algorithm can greatly minimize the interference by simultaneous transmission which aids the routing algorithm to divert traffic between nodes. The Channel assignment and routing problem can be combined together and optimized jointly to improve overall network performance. The impact of the interference can be controlled using routing metric and algorithm.

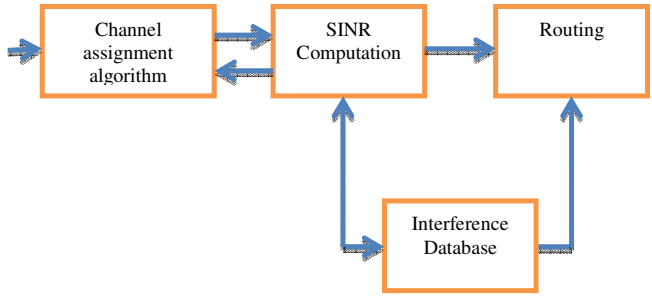

**Figure 6:** Block Diagram that represents the interaction between channel assignment algorithm and the SINR based routing algorithm.

**4.1 Channel Assignment Algorithm**

The efficiency of channel assignment primarily depends on the number of simultaneous transmissions in WMN and the number of simultaneous transmissions is limited by interference. The problem of reducing the signal interference and collision is represented as a graph coloring problem. Adjacent edges in the graph are assigned different colors and different channels will be allocated from 1 to 11.The physical graph of wireless mesh network in concern is represented as an undirected graph $G = (V,E)$; where V represents set of wireless routers, and E represents undirected edges between wireless mesh routers. We represent this channel assignment problem as an edge coloring problem of graph theory.

**Definition 1:** For an undirected graph G, conventional edge coloring algorithm assigns colors to the edges of G in such a way that no two adjacent edges are assigned with same color.

**Definition 2:** For an undirected graph G, strong edge coloring algorithm assigns colors to edges of G in such a way that no two edges in the neighborhood of utmost 2 hops is assigned with same color.

**Chromatic Index:** The chromatic Index (K) of the graph G is the minimum number of colors required to color the edges of a graph.

We assume static mesh routers each equipped with multiple radios. The transmission power of every mesh node is assumed to be the same. Since we use IEEE 802.11 b/g standard, the total number of available channels are 11. The mesh gateway node is assigned with the responsibility of assigning channel to the mesh routers, which first constructs a physical graph G and a set of active links A. If a link e connecting the nodes u and v, needs to be assigned to a channel c; it computes the total signal to noise plus interference ratio, with respect to the channel c for both the nodes. This process repeats for all 11 channels. Whichever channel provides the minimum value of interference will be chosen for the assignment. Initially it assigns channel (colors) to links connected to the mesh gateway. Each time, nodes at a particular distance are selected and the edges linked to them are given a color. In this paper color and channel are used interchangeably.

Vizing's theorem computes the number of colors in the edge-coloring problem of every undirected graph using at most one larger than the maximum degree d of the graph. Misra and Gries [10] describe an NP-complete algorithm for coloring edges of any graph with d + 1 colors, where d is the maximum degree of the graph.

The edge coloring algorithm assumes that there are N routers present in the network and it stores interference value in database for each link. Initially each link is assigned with zero which indicates the uncolored edge.

| Algorithm | Channel Assignment |
|---|---|
| **Input** | Let $G = (V, E)$ denote the network<br>V = Set of routers<br><br>$E \in V \times V$ is the set of undirected edges<br><br>Let $A = (V, E_A)$ Sub-graph of G selected by the algorithm |
| **Output** | Channels assigned to edges present in A |
| | 1. Let K = List of available channels (K=11)<br>2. Let $u_i$ = Root of the mesh network for i =1 to N<br>3. Let h= Number of edges incident on $u_i$ from A.<br><br>4. **For** all edges $e \in E_A$ **do**<br><br>5. Color(e) = 0<br>6. **while** count $\neq$ h+1<br>7. **For** i=1to N **do**<br>8. Assign color (A,G)<br>9. **end while** |

| Procedure | **Procedure** AssignColor (G1= (V1,E1) G2=(V2,E2))<br>1. **For** i = 1 to‖uncolored edges attached to $u_i$ in G1 ‖ **do**<br>2. Let p= uncolored edge attached to u<br>3. C1= Least interference channel not Used by links to G2.<br>4. c1 is selected based on signal- to-noise interference calculation<br>5. Assign c1 to link p<br>6. **if** such channel does not exist, **then**<br>7. Channel with minimum Interference is greedily assigned to link p.<br>8. **end if** |
|---|---|

Let $P_{x1}$ be the transmission power of node X1. Let $G_{x1y1}$ denote channel gain for nodes X1 and Y1, which depends on the distance between nodes X1 and Y1 and path loss index. Let $N_{y1}$ be the thermal noise at the receiver Y1.The SINR at receiver Y1 is given by

$$SINR_{x1y1} = \frac{P_{x1}G_{x1y1}}{N_{y1} + \sum_{z1 \neq (x1,y1)} P_{z1}G_{z1y1}} \qquad (5)$$

In this model, the available channel (color) is 11 means that K=11. For square topology (figure 10) if K = 4, then the edge coloring problem is NP-complete by Vizing's theorem. In random topology, nodes located close to each other, so nodes get more interference compared to square topology. Even though chromatic index for random topology is three, because of the interference as the algorithm uses 11 channels. The part of the random topology is selected to show the interference aware edge coloring in figure 7. The channel 9 assigned to link L1. So, the channel 9 cannot be given to links L2, L3, L4, L5 and L6 because they are within the interference range. For random topology (figure 9) if K=11, then strong edge coloring problem is deterministic and NP-complete by Vizing's theorem.

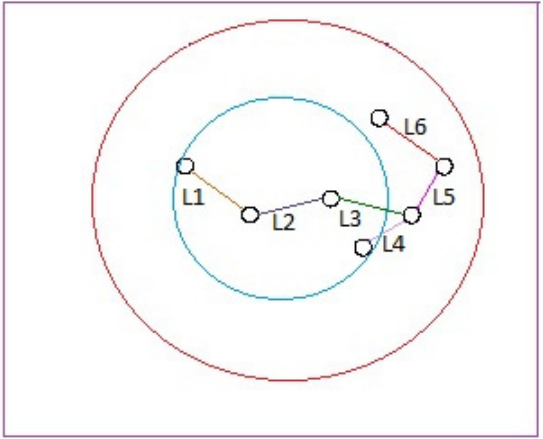

**Figure 7:** Interference aware edge coloring for random topology.

### 4.2 Routing based on SINR computation

In Traditional shortest path routing, the shortest route between the source and destination is determined based on the length of the path. The good routing protocol has to compute the shortest path which satisfies the following criteria: minimum path length, minimum end-to-end delay and maximum data rate. The good routing metric should be able to measure the link qualities and then helps the routing protocol in meeting the criteria. The design of a new routing metrics for wireless networks is challenging due to the fact that the wireless links are having the following unique characteristics:

**Packet loss:** If two communicating nodes are located with greater distance or any obstacle in the environment, then it leads to channel fading and increased loss ratio in links. Packet loss causes more delay and reduces the throughput.

**Packet transmission rate:** Based on the load and loss ratio on wireless link, Packet transmission rate may vary from time to time.

**Interference:** IEEE 802.11b/g standard operating in 2.4GHZ unlicensed spectrum may suffer from interference external to the wireless network such as microwave ovens and Bluetooth. Also, data transmission on one link may interfere with transmission in neighbouring links. Therefore, the routing metric should capture both inter-flow and intra-flow interferences, to choose the interference free path for routing packets.

The protocol interference model is used widely to obtain the interference information and it can be easily applied in theoretical analysis. But the protocol interference model is not perfect when compared with the physical interference model. The SINR is based on the physical interference model and it is particularly based on the exact transceiver design of systems. But, it is quite complex and difficult to apply in graph theory based algorithms. In spite of the complexity, the SINR model fetches good accuracy in interference calculation. Hence, the signal-to-noise plus interference ratio is proposed as a new routing metric to find interference on a wireless link. The SINR value is considered as link cost in routing algorithms. Our objective is to reduce the total cost of routing and at the same time to ensure that the total load on each wireless link is less than its capacity.

Using the channel assignment based on the graph coloring, the significant amount of reduction in the interference observed on simultaneous transmissions. In general, Routing from source (end-user) to destination (gateway) may take multiple hops, and the Mesh Gateway is connected to internet which provides broadband access to the source (end-user). The routing algorithm retrieves the topology from the channel assignment algorithm to accurately estimate the cost of the link. The SINR value of each link is calculated and stored into the database. The routing algorithm retrieves the SINR information from the database and the shortest route is constructed based on the least interference path.

**Problem Formulation**

The channel assignment algorithm is used as an input to the SINR based routing and the routing algorithm finds the best route from the user to any one of the gateway. A good routing protocol has to find an optimal path with in short span of time and it should also reduce the update frequency into the routing table to efficiently manage the network resources. It must be able to guarantee some level of quality of service. Note that, the considered grid topology consists of two gateways. If the value of SINR is high, then it indicates that there is a low interference in that link. The cost of each link is estimated from the SINR computation (Eq.6) and the capacity of each link is represented by $C_{x1y1}$. The capacity is calculated using Shannon's formula given in the equation 4. The flow of data from node x1 to its neighbour y1 over wireless link(x1,y1) is represented by $f_{x1y1}$.

$$SINR_{x1y1} = \frac{P_{x1}G_{x1y1}}{N_{y1} + \sum_{z1 \neq (x1,y1)} P_{z1}G_{z1y1}} \geq \beta \quad (6)$$

The channel gain $G_{x1y1}$ is estimated by a commonly used model $G_{x1y1} = d_{x1y1}^{-\alpha}$, Where $d_{x1y1}$ is the physical distance between nodes x1 and y1 and α is the path loss index. The SINR threshold is represented by β.

The cost of each link is estimated by interference experienced on that link as a result of simultaneous transmission and noise. The higher SINR value the better the quality of the link. Therefore, our objective is to maximize the SINR value and using the SINR based routing to deliver all the packets transmitted to gateway by the end user nodes, without exceeding the capacity of the link. The Routing problem is represented as follows:

$$\text{Maximize} \sum_{(x1, y1) \in E} SINR_{x1y1} \quad (7)$$

Subject to

$$\sum_{(x1,y1) \in E} f_{x1y1} - \sum_{(y1,x1) \in E} f_{y1x1} = d_{x1}, \ \forall\, x1 \in V \quad (8)$$

$$0 \leq f_{x1y1} \leq C_{x1y1} \quad (9)$$

$$f_{x1y1} \in Z^+ \quad (10)$$

Where $d_{x1}$ represents the data rate at which packets are produced at node x1 per seconds. The equation 8 ensures flow maintenance at each node. Second constraint (Eq. 9) ensures that the flow of data on a link should not exceed its channel capacity. Third constraint specifies that the packet flow is an integer value. When only the topology information is considered, the context is reduced to interference aware edge coloring problem. To provide fair and efficient solution, the second constraint compares the traffic on wireless link with available capacity. In Figure 8, algorithm of proposed SINR based routing is represented.

| Input | Network topology, active links and the SINR value from the database |
|---|---|
| Output | Shortest path from end users to WMN Gateway |
| Algorithm | Proposed routing algorithm |
| | 1. Read network configuration and set of active links from the channel assignment. 2. Set cost of all links to zero. 3. Read SINR value from the interference database. 4. Covert SINR value to the link cost and assign the cost to link. Minimum cost of any link is β. 5. Find the route to any one of the gateway (destination). 6. If more than one gateway nodes are available, then find the route with a least interference. i.e. Optimal route. 7. If not, then the route is established with gateway. |

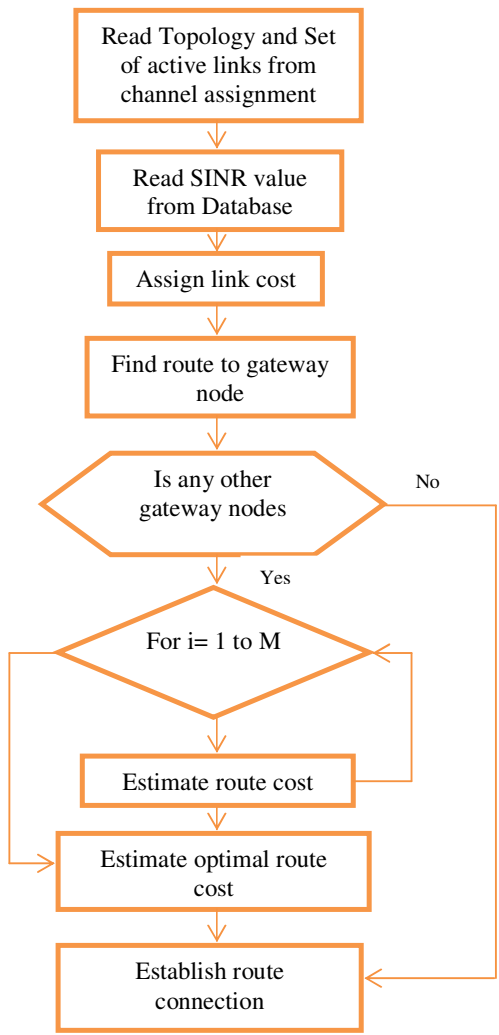

**Figure 8:** The proposed routing algorithm based on SINR

## 5. Simulation

### 5.1 Simulation Results: Channel assignment algorithm

NS2 based simulation is used to simulate the Interference Aware Edge Coloring problem. The NS2 version NS2.33 and, the patch for multi-channel multi-radio is included. Diverse types of topologies like square, random were used in the simulation, and the comparison of the orthogonal channel inputs with POC inputs were observed. The area dimension for our simulation is within 1000m × 1000m flat grid topology. The physical distance between two nodes is 200m in square topology; the transmission range is 250m for all the nodes and the interference range is 550m. But, the random topology distance between two nodes can differ randomly. Each node is equipped with multiple radios and data transmission rate is 11Mbps. The thermal noise power is set to -90 dB; the SINR threshold is set to -10 dB and the packet size is set to 1000bytes. The simulation was performed in 300s and the traffic types used in our simulation are Ftp and Cbr.

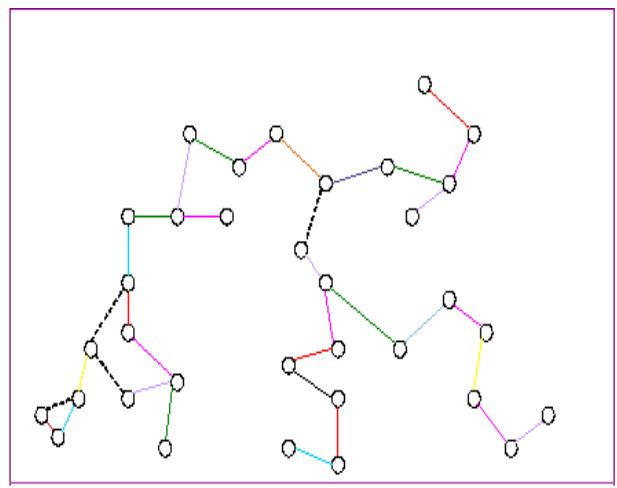

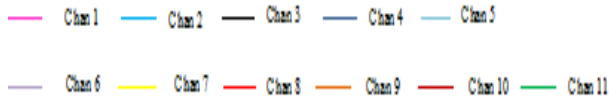

**Figure 9:** Random Topology

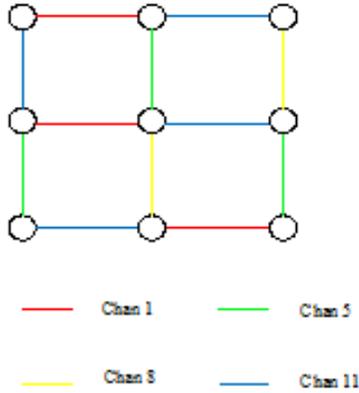

**Figure 10:** Square Topology

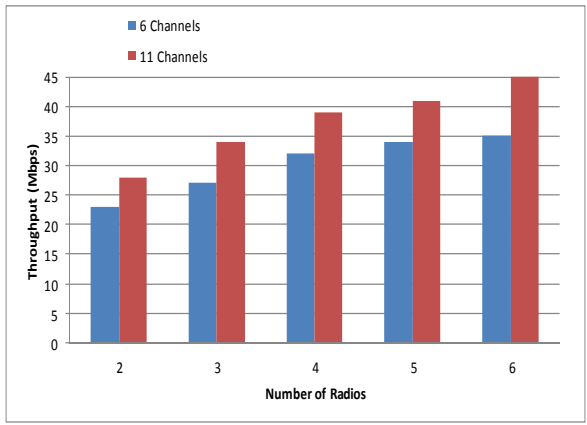

**Figure 11:** Throughput Vs Number of radios

To evaluate the channel assignment algorithm, the network throughput and the aggregate network capacity were evaluated. In Figure 11, the network throughput for the number of radios is compared and it has been observed that, the Interference Aware Edge Coloring channel assignment using 11 channels demonstrated the better throughput with spatial channel re-use. This shows that our method delivers the maximum throughput with more number of channels. When the number of radios and channels are increased, the network throughput also increased. Figure 12 shows, the aggregate network capacity for WMN. As the network capacity is increased dramatically after 5 channels, this clearly demonstrates that POC increases overall network performance in wireless mesh networks.

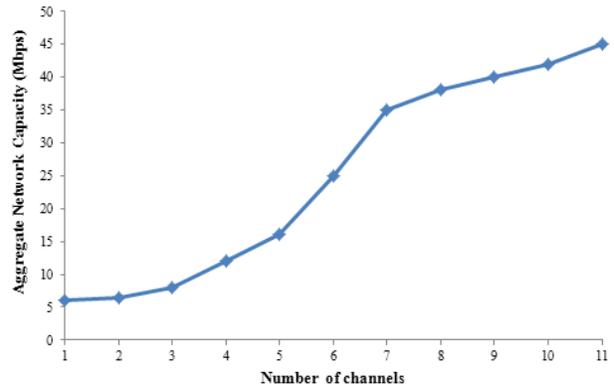

**Figure 12:** Aggregate Network capacity Vs. Number of channels

**5.2. Simulation results of Routing using SINR**

We have used grid topology consists of 16 nodes shown in figure 13 for interference aware routing. AODV routing algorithm is modified to use SINR value as a link metric instead of hop count to find the shortest route. When the source desires to transmit data to the destination but does not already contain the path to destination in its routing table, then it starts route discovery process. As part of the discovery process, the source sends route request packets (RREQ) throughout the network. The route request packet contains the link cost from the database, a RREQ identifier, the originator address, the originator sequence number, the destination address and the destination sequence number. The link cost contains the cost to travel from the source node to the next hop and it also contains the total cost that RREQ packet has traversed so far.

To uniquely identify a route request, the RREQ ID is combined with the source address. It ensures that RREQ packet is rebroadcasted only once, even though the node accepts the RREQ multiple times from its neighboring nodes. When the node receives the RREQ packet, it sends the route reply (RREP) back to the source node if it is the destination with sequence number equal to or greater than that contained in the RREQ. If the node has valid path to the destination, then it generates RREP packet to the source node. Otherwise, the node rebroadcasts the RREQ packet.

The source address and RREQ ID are verified to ensure that the RREQ packet has been received already. If it is received already, the packet is rejected. Upon reception of the RREP packet, the node will update or create its path to the destination. The link cost is updated and RREP packet will be forwarded to the source node. Finally, the source node will receive the RREP packet, if path exists from the source node to the destination. The data packets are transmitted to the destination on the discovered path.

When the link break happens, the upstream node propagates the RERR packets to the source node. Upon receiving the RERR, the source will reinitiate route discovery process.

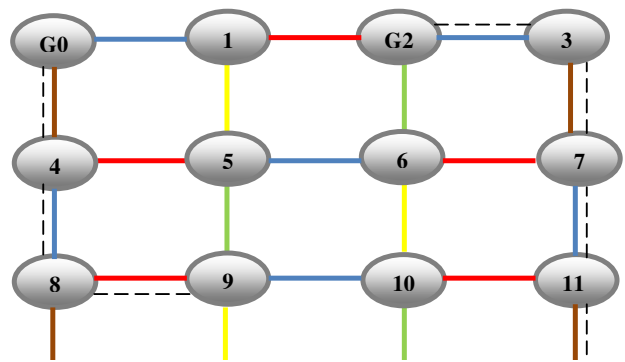

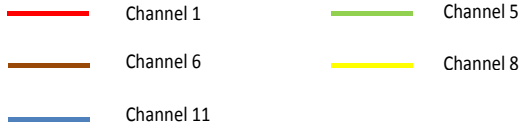

**Figure 13:** Shortest path from user to gateway node

We assume nodes 0 and 2 are gateways and the end-users are connected to router 14 and 9, the routing algorithm calculated shortest path as shown in figure 13. The routing algorithm finds route 9-8-4-G0 and 14-15-11-7-3-G2 as optimal and less interference path based on the SINR value. The shortest path 14-10-6-G2 based on the minimum hop count experiences higher interference level, packet drops may happen compared to the optimal path 14-15-11-7-3-G2.

Nodes 1, 5, 6, 9 and 14 are assumed as sources and optimal path to the Gateway is given by:
    Source 1: 1– G0
    Source 2: 6 – G2
    Source 3: 5 – 4 – G0
    Source 4: 9 – 8 – 4 – G0
    Source 5: 14 – 16 – 11 – 7 – 3 – G2

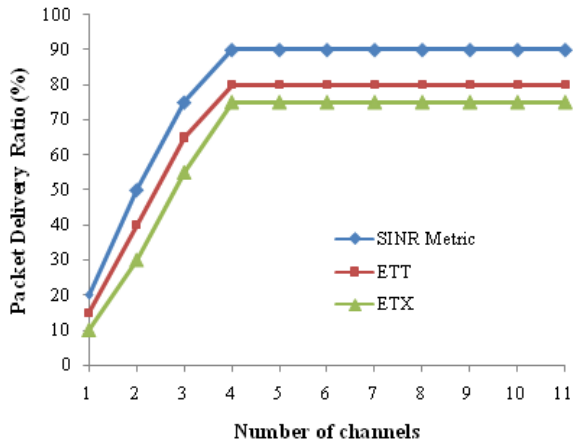

**Figure 14:** Packet Delivery Ratio versus Number of Channels

The performance of routing algorithm is compared with the ETX and ETT metrics as benchmarks. We evaluate the following metrics, to verify the performance of SINR routing:
    1) Packet delivery ratio
    2) End-to-End Delay
    3) Routing Overhead

The simulation results of routing algorithm is shown in figure 14, 15 and 16. Packet delivery ratio is estimated by counting the number of packets delivered successfully to the destination. In figure 14, the packet delivery ratio for number of channels is compared. It must be noted that the packet delivery ratio is increasing as the number of channel increases and maintains the same PDR after 4 channels. 90% of the packets delivered to the destination when the SINR metric is used in Multi-Channel WMN. End-to-End delay is the average delay it takes for a data packet to travel from the source to the WMN gateway.

Figure 15 shows how end-to end delay changes against the number of channels. It includes the delay made by the route discovery process and the queue delay. The routing overhead generated by our routing algorithm is shown in figure 16. It shows how many times the packets are retransmitted due to the interference on routing path. The routing overhead is estimated by number of retransmission needed per connection between an end user and WMN Gateway. The simulation results in Figure 16 show that our method requires lesser retransmissions compared to other two metrics. Hence, the shortest path estimated using our SINR interference method is more reliable.

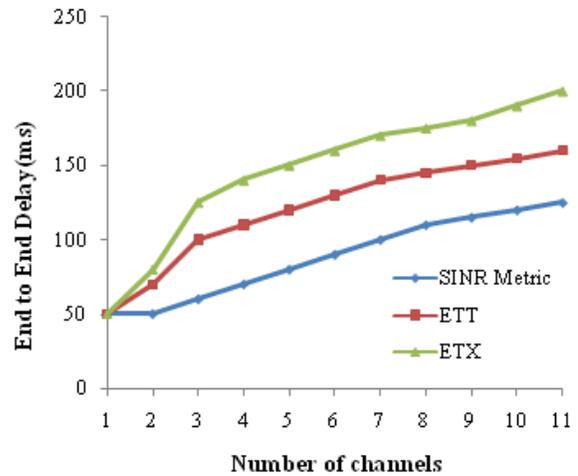

**Figure 15:** End-to End delay versus Number of Channels

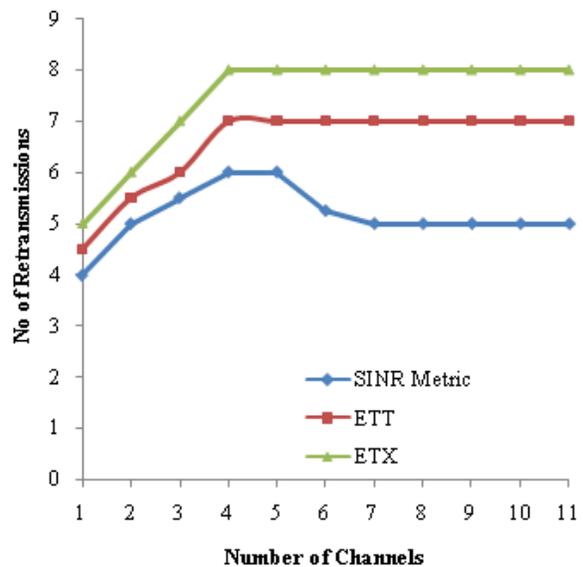

**Figure 16:** Routing Overhead

## 6. Conclusion

This work presents joint channel assignment and routing in Wireless Mesh Networks which uses the POC in addition to the orthogonal channels and a new routing metric called signal-to-noise plus interference ratio (SINR) value. We considered the channel capacity constraints for each traffic flow. From the simulation, we conclude that, Partially Overlapped Channels can improve the overall performance of the network. Moreover, the comparison with ETX and ETT shows the performance and the effectiveness SINR metric and the proposed method. The frequency band of IEEE802.11b/g is to be completely utilized. This method supports more number of parallel transmissions. In future, we are planning for the following cross layer framework:

1) Physical/Transport Cross layer design to adjust the transmission rate of the source to avoid congestion in the network. 2) Network/Transport Cross layer design to support traffic engineering.